\documentclass[12pt]{iopart}
\usepackage{iopams}
\usepackage[latin1]{inputenc}
\usepackage{graphicx}
\begin{document}

\title[]{Towards a full Heusler alloy showing room temperature half-metallicity at the surface}

\author{M Cinchetti$^1$, J-P Wüstenberg$^1$, M S\'{a}nchez Albaneda$^1$, F Steeb$^1$, A Conca$^2$, M Jourdan$^2$, M Aeschlimann$^1$}
\address{$^1$University of Kaiserslautern, Institute of Physics, Erwin-Schrödingerstr. 46, 67663 Kaiserslautern, Germany}
\address{$^2$University of Mainz, Institute of Physics, Staudinger Weg 7, 55128
Mainz, Germany } \ead{cinchett@rhrk.uni-kl.de}
\begin{abstract}
In this article we investigate the surface spin polarization in a
100\,nm Co$_2$Cr$_{0.6}$Fe$_{0.4}$Al (CCFA) film grown {\it ex
situ} epitaxially on MgO(100) with a 10\,nm Fe buffer layer by
means of spin resolved photoemission. We show that a careful {\it
in situ} preparation of the sample surface leads to values for the
room temperature spin polarization up to 45\% at the Fermi level.
To our knowledge, this is the highest value measured  so far at
the surface region of a full Heusler alloy at room temperature.
\end{abstract}

\submitto{\JPD}
\newpage
\section{Introduction}
The success of modern spintronics devices, i.e. devices relying on
the electron spin as carrier of information depends crucially on
the ability to store, transport and manipulate the spin state of
an electron within a properly chosen material system. While
storage is traditionally accomplished by the use of
magnetoresistive effects in magnetic multilayer systems, the
latter problems seem to be trackable by the use of (partly)
semiconductive materials, where typical spin diffusion lengths are
much larger and material properties can be tuned precisely to
establish a coupling to external fields for coherent spin
manipulation purposes \cite{zutic04}. However, the conductivity
mismatch \cite{Schmidt00,Attema06} between ferromagnetic metals
and semiconductors which hinders efficient spin injection is still
an unsolved issue. There are two possible approaches to overcome
this issue \cite{Wolf01}: realize pure semiconductor devices (for
which ferromagnetic semiconductors are needed) or search for new
materials exhibiting large carrier spin polarization. A promising
class of materials for the latter possibility consists of half
metallic Heusler alloys, exhibiting metallic behavior for one spin
direction and a band gap for the other. This should lead to a full
spin polarization $P$ at the Fermi level $E_F$, which is more
generally defined as the ratio of spin up and spin down electron
occupation numbers with respect to a certain energy and given
quantization axis:
\begin{equation}\label{Polarization_formula}
P(E)=\frac{N_{up}(E)-N_{down}(E)}{N_{up}(E)+N_{down}(E)}
\end{equation}

In order to be attractive for technical applications, a Heusler
alloy should satisfy  in principle three requirements: such a
material should show a full polarization: (i) at $E_F$ (i.e., be a
half-metal), (ii) not only in bulk but up to the very surface
region, and (iii) at room temperature.

The fabrication quality of such alloys is improving steadily. This
is shown for example in the case of Co$_2$FeSi by the increasing
coherence between measured and predicted values for the
element-specific number of magnetic moments $\mu_B$ per unit cell
\cite{Kandpal06,Wurmehl05}. However, the connection to the spin
polarization at $E_F$ as the relevant parameter is not
straightforward and relies heavily on the chosen model potential
in calculations.

To our knowledge, a Heusler compound satisfying all conditions
(i),(ii) and (iii) has not yet been reported. Half metallicity has
been experimentally verified in NiMnSb but only for the bulk
material and at $T=27$\,K \cite{Hanssen90}. On the other hand,
surface sensitive techniques failed to demonstrate
half-metallicity of such compound. It is not straightforward that
the surface region of a Heusler alloy will show the same electron
spin polarization (ESP) as the bulk material. In principle,
extrinsic as well as intrinsic mechanism may reduce the spin
polarization at the surface. Among the extrinsic mechanisms,
chemical and magnetic disorder have been discussed to play a
relevant role \cite{Raphael02,Tokunaga01}. Intrinsic mechanisms,
that take place also at perfectly ordered surfaces, are for
example magnetic fluctuations and modifications in the surface
band structure with respect to the bulk one. In a recent
publication, Kolev \textit{et al}. \cite{Kolev05} have
investigated the (100) surface of NiMnSb using spin-resolved
appearance potential spectroscopy and found a significantly
reduced spin asymmetry with respect to calculations based on the
bulk electronic structure. By a careful analysis, the authors
could exclude chemical disorder, structural defects at the
surface, as well as overall stoichiometric disorder as responsible
mechanisms for the observed reduction of spin polarization. In
another paper Wang \textit{et al}.\ \cite{Wang05} have studied
single-crystalline Co$_2$MnSi films and found for $P(E_F)$ a
maximal value of 12$\%$ measured with spin-resolved photoemission
at room temperature. They attribute this high discrepancy with the
expected value of $100\%$ to partial chemical disorder in the
Co$_2$MnSi lattice.

Co$_2$Cr$_{0.6}$Fe$_{0.4}$Al (CCFA) has been intensively studied
both experimentally
\cite{elmers03,auth03,felser03,kelekar05,hirohata05,conca06,block06}
and theoretically \cite{miura04,antonov05,Wurmehl06}. The interest
in this compound is due to its high Curie temperature of 760\,K
and its theoretically predicted high value of volume magnetization
of $3.8\,\mu_B$ per formula unit \cite{Galanakis02}. To our
knowledge, the spin-resolved photoemission results presented here
for thin films of such full Heusler compound set the actual room
temperature ESP record for Heusler alloys to 45\% at the Fermi
level and at the surface region, a value reflecting the increasing
sample quality as well as more and more sophisticated surface
preparation methods. This constitutes a very important step
towards the goal of producing an alloy satisfying simultaneously
conditions (i), (ii) and (iii).

\section{Experimental setup}

The thin films samples are prepared in a commercial MBE deposition
system with analysis chamber which is complemented by a home build
sputter chamber.

The spin-resolved photoemission experiments are performed in a
separate UHV system with a base pressure of $<10^{-10}$\,mbar,
equipped with an Ar$^+$ sputter gun. Surface characterization is
achieved by LEED as well as Auger analysis and by identifying
characteristic features of the photoemission spectrum. The latter
is obtained by irradiating the sample with the s-polarized 4th
harmonic (photon energy $5.9$\,eV) of a 100\,fs Ti:Sapphire
oscillator (Spectra Physics Tsunami). The laser light angle of
incidence is set by geometry to $45\,°$, while the spectra are
taken in normal electron emission. Spin resolved photoemission
spectra are recorded by a commercial cylindrical sector analyzer
(Focus CSA 300) equipped with an additional spin detector based on
spin-polarized low-energy electron diffraction (Focus SPLEED). The
achieved energy resolution is 150\,meV, the acceptance angle of
the analyzer is $\pm 13^{\circ}$.  For the measurements the films
are remanently magnetized by an external in-plane magnetic field.

\section{Sample preparation and pre-characterization}
The thin films of Co$_2$Cr$_{0.6}$Fe$_{0.4}$Al were deposited by
dc magnetron sputtering from a stoichiometric target. Epitaxial
growth in (100) orientation was obtained on MgO(100) substrates
employing an Fe buffer layer (10\,nm thickness). X-ray diffraction
analysis revealed that the films grow in the B2 structure. The
sample surface after deposition and annealing is well ordered as
demonstrated by electron diffraction (RHEED)\cite{Conca06b}. To
protect the surface of the 100\,nm thick Heusler-film from
oxidation during the transport between the preparation and the
photoemission chamber, the sample was capped  by 4\,nm of Al.\\
Before characterization with spin resolved photoemission, the
magnetic properties of the sample were studied \textit{ex situ}
using magneto-optical Kerr effect. A fourfold in-plane anisotropy
was reported \cite{Hamrle06}, reflecting the crystallographic
symmetry of the CCFA film.

\section{In situ sample treatment} \label{Preparation}
In order to obtain reproducible results with spin resolved
photoemission, the (100) surface of the the sample has to be
carefully prepared {\it in situ}. Since this experimental
technique is sensitive to the surface region \cite{Pit06},  the
protecting Al cap-layer has to be carefully removed. A further
reason requiring a careful preparation is that the crystal
structure of CCFA easily allows for atomic disorder, either by
interchange of atoms or by partial occupancy \cite{elmers03}.

Due to the critical role played by the preparation procedure, we
will describe it in detail.

\begin{itemize}
\item[(A)] As first step the Al cap-layer was removed from the
sample by  several cycles of Ar$^+$ ion sputtering at moderate
rates (sputtering energy $500$\,eV at an angle of incidence of
$\pm 65\,°$ with respect to the surface normal). The successful
removal of the cap layer was checked by means of Auger
spectroscopy and by monitoring the changes in the photoemission
spectrum.

\item[(B)] The sample was heated at $300\,°$C for 12\,hours before
the first characterization attempts.

\item[(C)] Without further treatment, the prepared surface shows
no LEED patterns, and the measured polarization is identically
zero over the whole spectral range of about $2$\,eV below $E_F$.

\item[(D)] As a next step the sample was treated with further
sputtering and heating (at $300\,°$C) cycles until a clear LEED
pattern was observable.

\item[(E)] For the last cycles the sample temperature was held at
$300\,°$C during Ar$^+$ ion sputtering. After sputtering the
sample was heat flashed for approx.\ 1\,min.\ to $450\,°$C. The
last step was repeated before each measurement, and assures the
reproducibility of the obtained results.
\end{itemize}

\section{Experimental results and discussion}

\begin{figure}[t!]
\begin{center}
\includegraphics[height=4 cm,keepaspectratio=true]{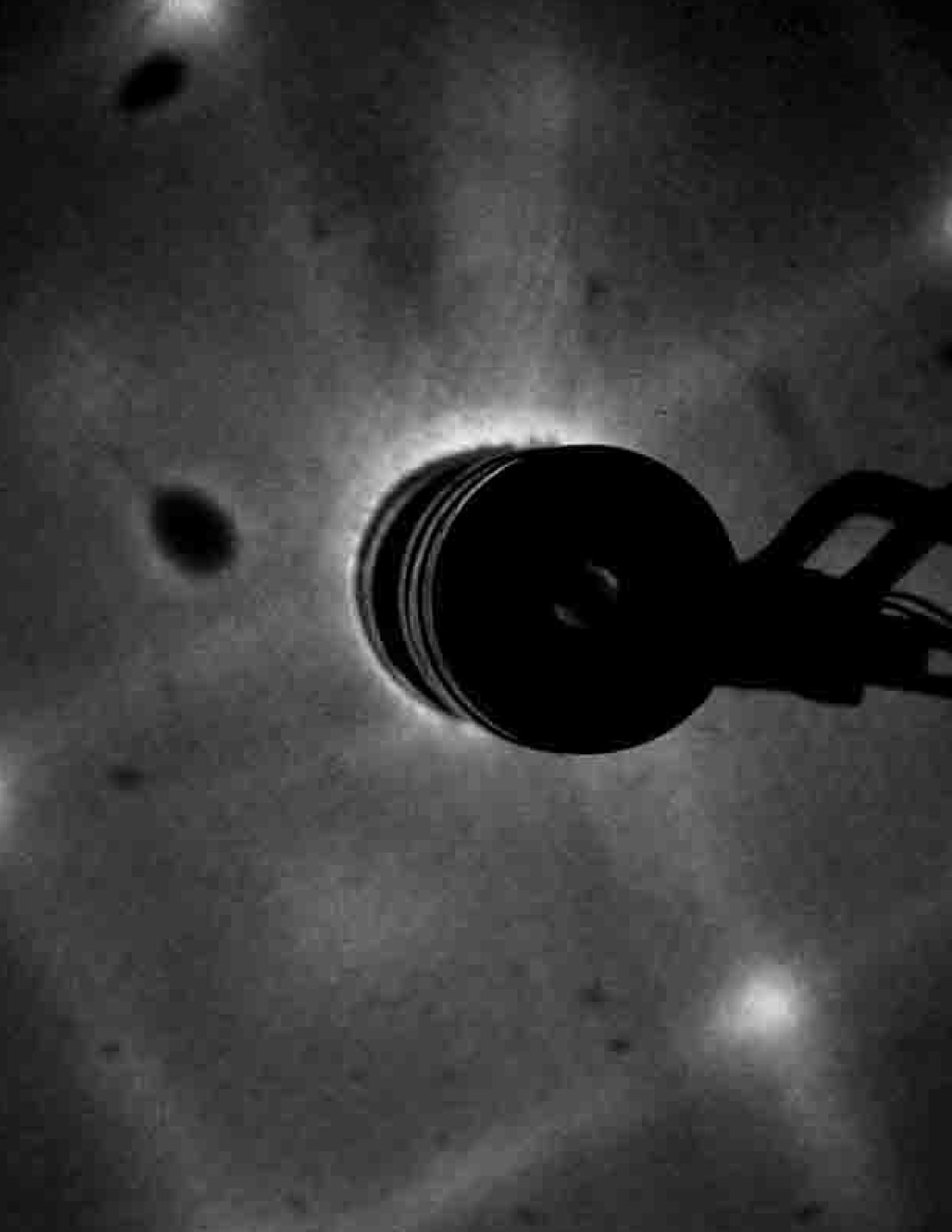}
\includegraphics[height=4 cm,keepaspectratio=true]{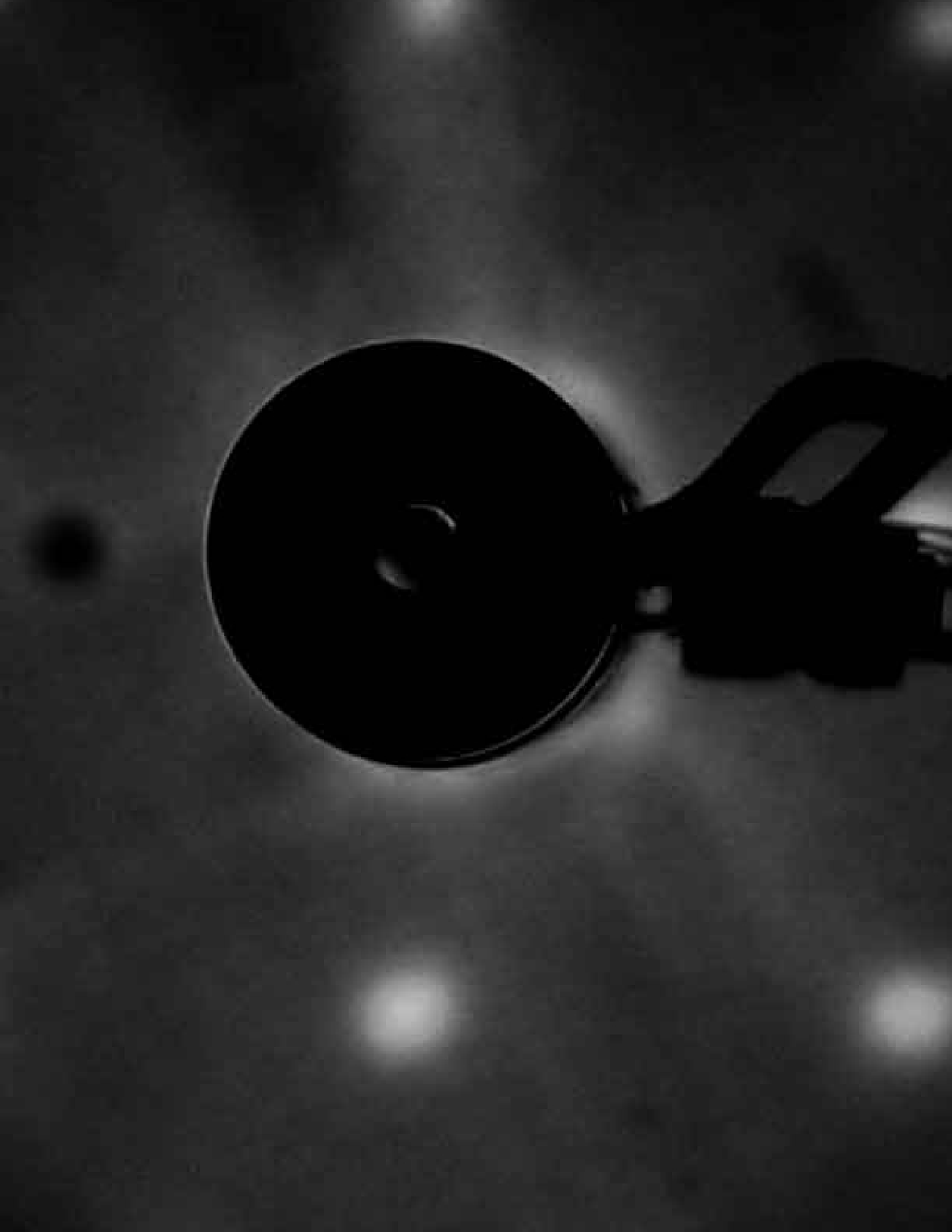}
\caption{(a) LEED pattern of the sample after the steps (A) to (D)
of the {\it in situ} preparation procedure, taken at 88\,eV
primary electron energy.\\ (b) LEED pattern of the sample after
the full  {\it in situ} preparation procedure (steps (A) to (E)) ,
taken at 100\,eV primary electron energy.} \label{LEED}
\end{center}
\end{figure}
Figure \ref{LEED} (a) shows a LEED image obtained from the sample
after the steps (A) to (D) of the {\it in situ} preparation
procedure described in Section \ref{Preparation}. The picture was
recorded at 88\,eV  primary electron energy and demonstrates a
clear fourfold symmetry in correspondence with the cubic bulk
lattice. If step (E) is also applied, then the LEED pattern
obtained from the sample is shown in Figure \ref{LEED} (b). The
picture was recorded at 100\,eV primary electron energy. Compared
to Figure \ref{LEED} (a), Figure \ref{LEED} (b) shows a clearer
LEED pattern, demonstrating that the preparation step (E) leads to
a better rearrangement of the surface atoms.

\begin{figure}[h!]
\begin{center}
\includegraphics[width=10 cm,keepaspectratio=true]{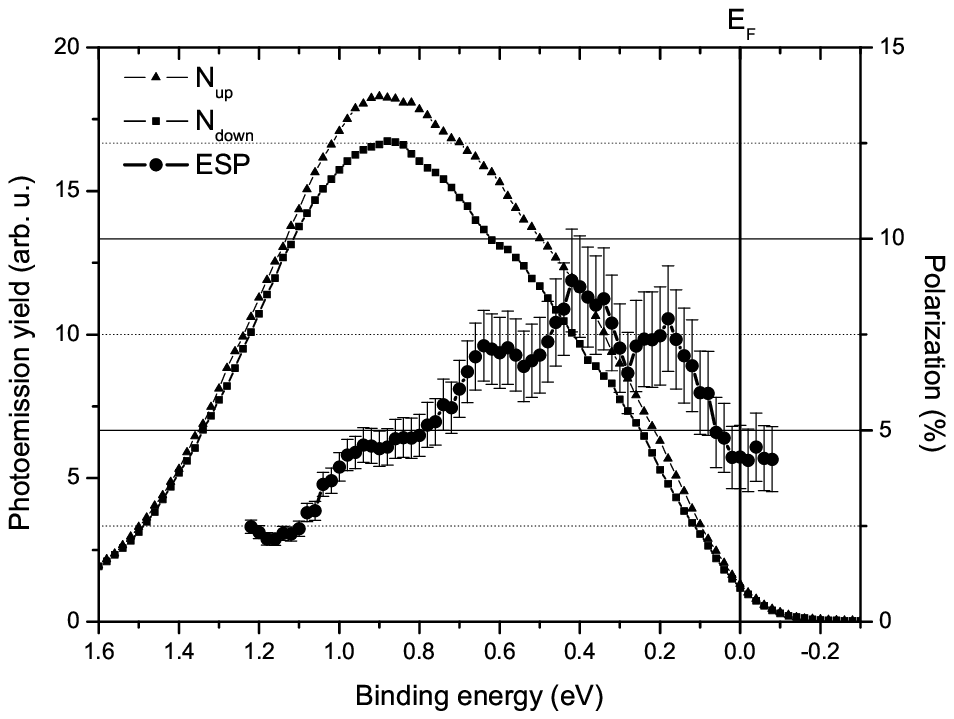}
\caption{Spin resolved photoemission spectra and ESP of the CCFA
thin film after preparation steps  (A) to (D).\\
Left scale: photoemission yield from majority electrons ($N_{up})$
with filled triangles and minority electrons ($N_{down}$) with
filled squares.\\ Right scale: ESP as obtained from Equation
(\ref{Polarization_formula}), with filled circles.}
\label{Spectrum_before}
\end{center}
\end{figure}
Figure \ref{Spectrum_before} shows on the left scale the spin
resolved photoemission spectra recorded from CCFA after after the
steps (A) to (D) of the {\it in situ} preparation procedure
described in Section \ref{Preparation}, i.e.\ corresponding to the
LEED pattern of Figure \ref{LEED} (a). The spectra for the
majority electrons ($N_{up}$) are represented with filled
triangles, the one for minority electrons ($N_{down}$) with filled
squares. They show the Fermi level $E_F$ (showing up at 0\,eV in
the chosen binding energy scale) and a monotonous increase of the
photoemission yield up to  $\sim 0.9$\,eV. The low energy cutoff
is approximately at $1.2$\,eV. The ESP, calculated from those two
curves using Equation (\ref{Polarization_formula}), is shown in
the same picture on the right scale (filled circles). Overall, the
ESP does not reach values higher than 10\%. In particular, it
starts from $2.5$\% at the low energy cutoff, and shows three
peaks at $0.6$\,eV, $0.4$\,eV and $0.2$\,eV. Then, it decreases
again towards $E_F$ to reach $P(E_F)\sim 5$\%, well below the
expected value of 100\%.

\begin{figure}[h!]
\begin{center}
\includegraphics[width=10 cm,keepaspectratio=true]{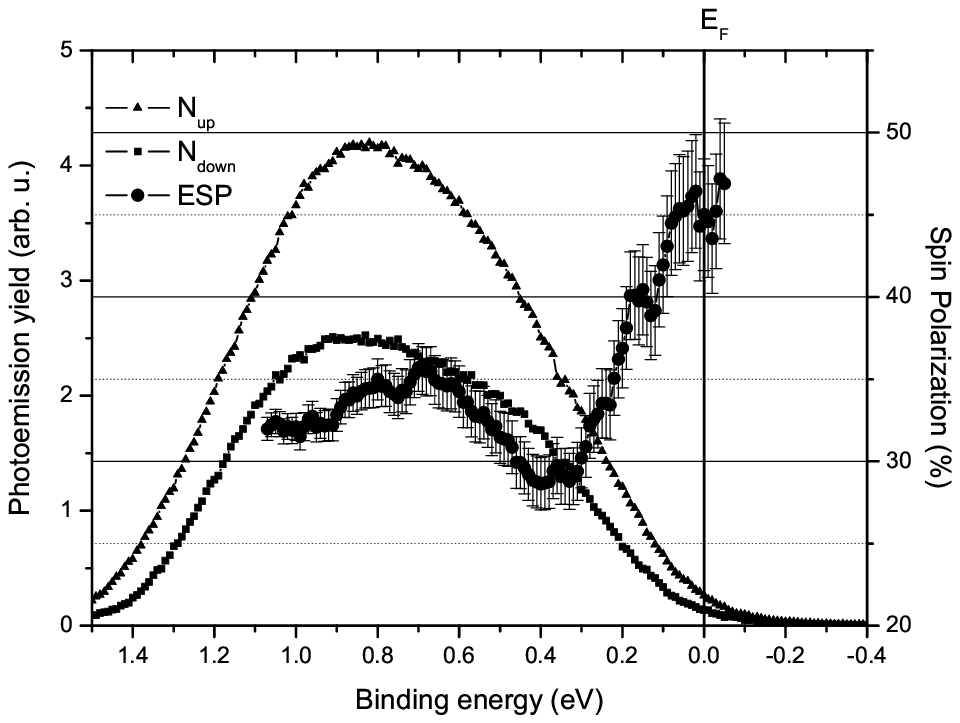}
\caption{Spin resolved photoemission spectra and ESP of the CCFA
thin film after the full preparation procedure (steps  (A) to (E)).\\
Left scale: photoemission yield from majority electrons
($N_{up})$) with filled triangles and minority electrons
($N_{down}$) with filled squares.\\ Right scale: ESP as obtained
from Equation (\ref{Polarization_formula}), with filled circles.}
\label{Spectrum_after}
\end{center}
\end{figure}
Figure \ref{Spectrum_after} shows on the left scale the spin
resolved photoemission spectra recorded from CCFA after the full
{\it in situ} preparation procedure (steps (A) to (E)), i.e.\
corresponding to the LEED pattern of Figure \ref{LEED} (b). Again,
the spectra for the majority electrons
 are represented with filled triangles, the one for
minority electrons  with filled squares. In this case, the two
spectra have remarkably different intensities, resulting in the
ESP depicted with filled circles on the right scale. Between the
low energy cutoff ($\sim 1.1$\,eV) and $0.4$\,eV binding energy,
the ESP assumes values between approximately 35\% and 30\%, then
it increases more or less steadily and reaches the value
$P(E_F)\sim 45$\%. This value is to our knowledge the highest spin
polarization at the surface region reported so far by means of
spin resolved photoemission from a full Heusler alloy at room
temperature. The remarkable difference in the absolute value and
energy dependence of the ESP in Figure \ref{Spectrum_before} and
Figure \ref{Spectrum_after} is a clear indication for the extreme
importance of the surface preparation procedure prior to the
measurements. Indeed, Correa {\it et al.}\ \cite{Correa06} have
reported for the half Heusler alloy NiMnSb(100), that an optimized
annealing process  after sputtering is essential to restore the
bulk stoichiometry of the surface, disturbed by the previous
sputtering. Evidence for such behavior was taken from the analysis
of normal-emission
ultraviolet photoemission spectra.\\
The specific surface treatment employed in our paper is the result
of an empirical optimization of the surface preparation prior to
the measurements, aimed to reach the highest possible value of the
surface spin polarization. The achieved results show that for this
particular Heusler alloy, the reduction of the surface spin
polarization can be -at least partially- circumvented by selecting
an adequate preparation procedure. This is of course very
important also for the production of spintronics devices, where
surface and interface between different materials play a crucial
role for the injection of spin polarization.

 The increasing of the ESP towards $E_F$ in Figure \ref{Spectrum_after}
 is a clear indication for the presence of a band gap (at least partial) for the minority
electrons, as expected from theoretical calculations
\cite{Wurmehl06}. However, at this point we have to notice that
photoemission from a single crystal gives information only about
direct transitions taking place near the surface Brillouin zone
center. Thus, the comparison of our experimental data with density
of states (DOS) calculations, as the one reported in
\cite{Wurmehl06}, can be only qualitative. Further angle- and
spin-resolved measurements are planned in the next future.
Together with  band structure calculations (including correlations
effects) this will give further precious information about the
electronic band structure of the CCFA Heusler alloy.

\section{Conclusions}
In this paper we have found an optimized  preparation procedure
for the (100) surface of the full Heusler alloy
Co$_2$Cr$_{0.6}$Fe$_{0.4}$Al.  Applying such procedure, we have
obtained a room temperature surface spin polarization of 45\% at
the Fermi level by means of spin resolved photoemission. To our
knowledge, this is the highest value reported so far by means of a
surface sensitive experimental technique. We proposed that
implementing such a procedure could lead to a better performance
of spintronics devices based on Heusler alloys.

\section*{Acknowledgements}
These studies were funded by the DFG Forschergruppe FOR 559/1 "New
Materials with High Spin Polarization"

\end{document}